# Planetary statistics and forecasting for solar flares

Eleni Petrakou[*(1)], Iasonas Topsis Giotis[†(2)]

*(1) Athens, Greece*
*(2) Ernst & Young Brussels, Belgium*

*18 June 2020*

**Abstract:** *Indications are presented for a significant connection between the relative motion of the planets and the appearance of energetic solar flares. Based on the records of the last four decades, the analysis highlights remarkable features and a lack of randomness in the data. The indications are supported further by the predictive power of a preliminary application to forecasting with machine learning methods.*

## 1. Introduction

Recent studies have indicated a connection between long-term solar activity and the relative motion of the planets Jupiter and Saturn[1], and between short-term solar activity and the positions of the inner planets[2]. Although a planetary role in solar activity has often been investigated before, what differentiates these studies from previous ones is the analysis of solar flares, instead of the commonly used sunspots, and the use of statistical analysis, instead of frequency analysis. (Two older studies which focused on statistical analysis of planetary positions for solar effects are [3] and [4].)

Specifically in [1], there were indications that the relative heliocentric motion of Jupiter and Saturn triggers the long-term solar activity. In the present article this concept is expanded by examining the role of the relative motion of more planets in the appearance of individual energetic solar flares.

Although no proposal is made about an underlying physical mechanism yet, it is reasonable that if such a mechanism involves the two gas giants then it applies to the inner planets as well. Additionally, it can be thought that the slower motion of the massive gas giants would result in larger and long-term effects, i.e. modulation of the solar cycle; while the faster planets would be associated with more variable and short-term effects, probably individual flares. It can be added that even though planetary modulation of the solar activity might sound unlikely, in essence it is one of the few permanent sources of perturbation on the Sun.

---

* e-mail: eleni@petrakou.net
† e-mail: jtopsis@gmail.com

Section 2 will present a number of indications for a relation between the appearance of energetic flares and the relative motion of the planets and Section 3 will discuss briefly the special case of alignment of many planets. Section 4 presents a preliminary application of this relation in the forecasting of solar flares, using machine learning techniques. Section 5 concludes with a short discussion. Parts of this work have been previously presented in [5].

## 2. Solar flares and relative planetary angles

For all pairs formed by the five innermost planets, their relative motion seems to be strongly related to the presence of solar flares. In this Section qualitative and quantitative indications are presented in favor of this relation being non-random.

In the following, "flares" will refer to all solar flares of X-ray flux intensity classes M and X in the years 1977-2019, the time range for which continuous records exist. "Angle" refers to the heliocentric ecliptic longitude. The flare records comprise the measurements of the NOAA SMS and GOES satellites[6]; sunspot records are provided by the Royal Observatory of Belgium[7], and planetary positions are calculated with NASA's HelioWeb tool[8]. The analysis was performed with the ROOT toolkit[9].

By "planets", in the following we refer to the five innermost planets. Saturn and especially Uranus and Neptune are not included because of their slow motion, which makes the variation of their position relative to the innermost planets meaningless on short timescales.

A first indication is that the distributions of flare counts as a function of the relative planetary angles point strongly to the absence of randomness. As an example Fig.1a shows the case for the angle between Earth and Venus, but all ten relative angles have similarly "jagged" appearance. Furthermore, the effect seems to be amplified when using the absolute value of the relative angles, as in Fig.1b. The absolute values are used in the following plots unless otherwise mentioned.

These distributions can be compared in terms of randomness to the distributions of days without any flares. Instead of the full sample with an arbitrary number of flares per day, only one entry per day will be used for flares, in order to have a more accurate comparison. The resulting distribution for the days with flares as a function of the angle between Earth and Venus is shown in Fig.2a, and the distribution for the absence of flares in Fig.2b.

A straightforward test of randomness consists of the goodness-of-fit for fitting the data with a straight line parallel to the x-axis, thus testing for uniformity. For the plots in Fig.2, each comprising 84 points, the fit results in a $\chi^2$ value of 167 for the presence of flares and 43 for the absence of flares. Therefore, a hypothesis of randomness can be rejected at a level of significance much greater than 99% for the days with flares, while it is strongly supported for the days without flares. There is very little change in the $\chi^2$ values when fitting with higher-order polynomials, and specifically with 3$^{rd}$-order polynomials in order to account for the effect of planetary eccentricities ($\chi^2$ values of 160 and 34 respectively).

The fits result in a difference of one order of magnitude in the $\chi^2$ values between the two cases, and the same conclusion, for all ten relative planetary angles.

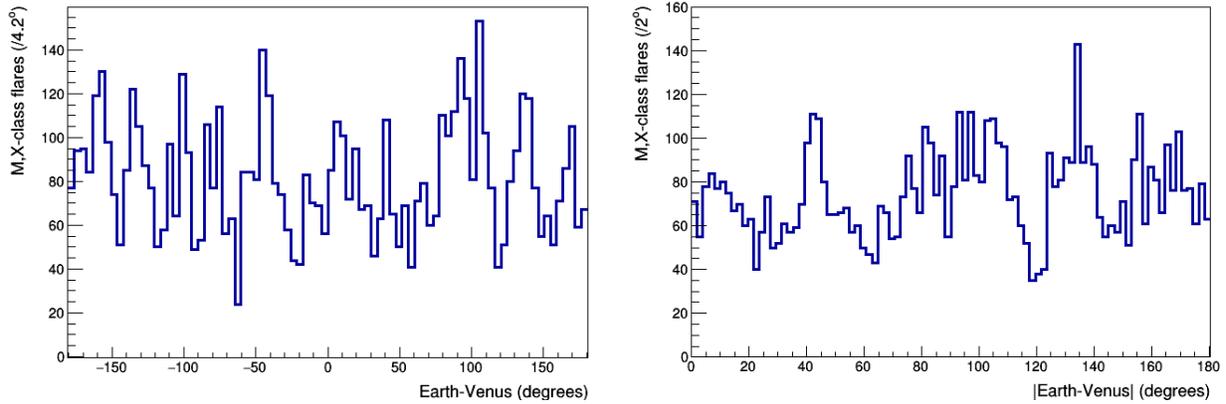

Figure 1. (a) Counts of M,X-class flares as a function of the relative angle between Earth and Venus in the years 1977-2019. (b) Same, as a function of the absolute angle.

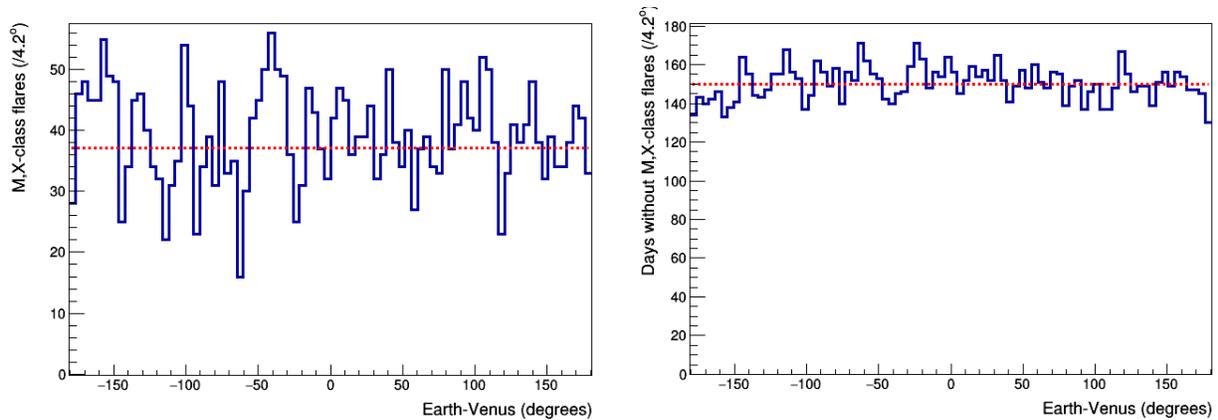

Figure 2. (a) Counts of days with M,X-class flares as a function of the relative angle between Earth and Venus in the years 1977-2019, and a fitted line parallel to the x-axis, representing a uniform distribution. (b) Same, for the days without M,X-class flares.

In certain cases "intuitive rules" seem to appear in these distributions. A pronounced one is a decrease in activity around 90º in the angle between Jupiter and Earth, shown in Fig.3.

The records cover 39 heliocentric synods between the two planets. However, a possible objection can be that the number of events in only a few years, happening to be less active than the rest, could amplify the overall effect of a statistical fluctuation. In Fig.4, though, it is shown that this decrease around 90° is present in both cycles 21 and 22, and probably also 24.

It can be noted here that the decrease in activity around 90° was found to be significant in the case of the relative angle between Saturn and Jupiter and their possible role in the solar cycle[1].

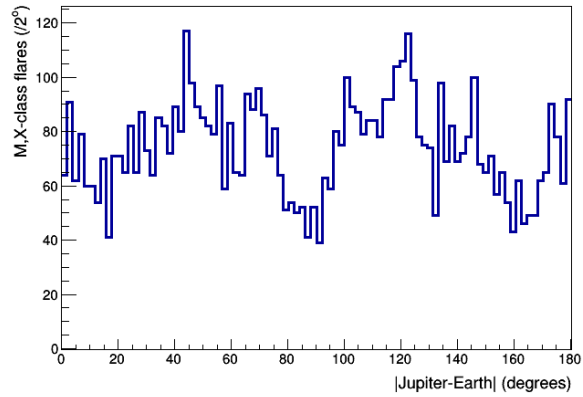

Figure 3. Counts of M,X-class flares as a function of the absolute relative angle between Jupiter and Earth in the years 1977-2019.

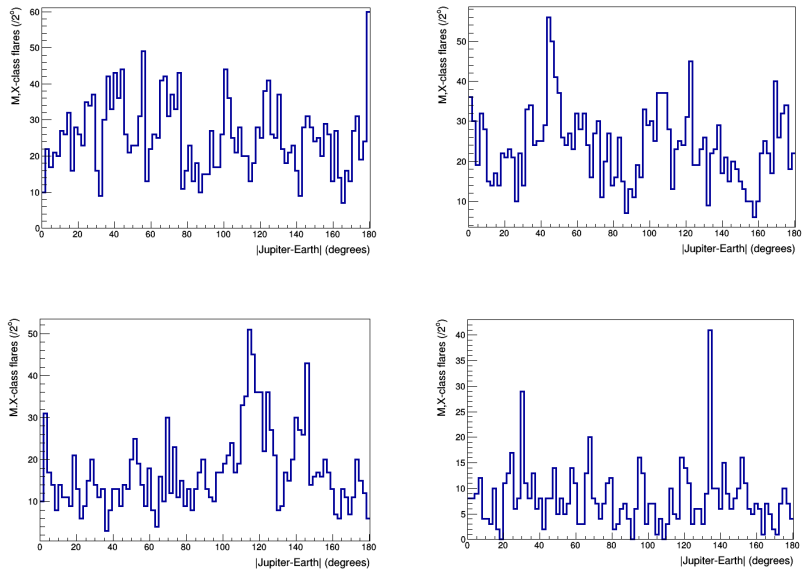

Figure 4. Counts of M,X-class flares as a function of the absolute relative angle between Jupiter and Earth in the years 1977-2019, individually for each solar cycle. (Upper left: cycle 21, upper right: 22, lower left: 23, lower right: 24.)

Furthermore, indications exist for a similar effect in the combinations of more than two planets, as is reasonably expected if the above hold.

As an example, Fig.5a shows the distribution of flare counts as a function of the angle between Mars and Earth, plotted separately for two "windows" of the angle between Jupiter and Earth. The windows cover ±5º around 120º and 90º, which correspond to the overall highest and lowest activity, as shown in Fig.3. The distribution is seen to form specific features, different for each window (to some extent these arise from individual lengths of increased activity, but not exclusively). In addition, this structure persists when examining only the X-class flares, as shown in Fig.5b.

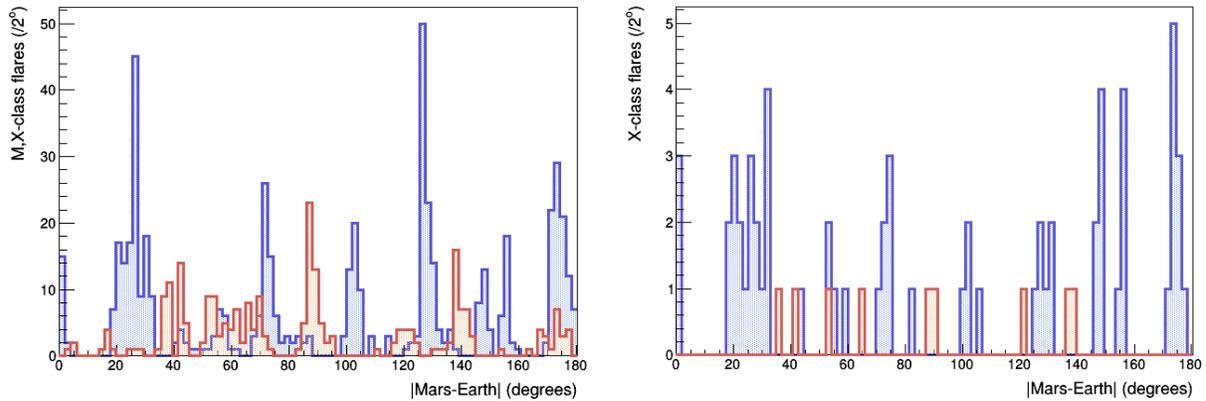

Figure 5. (a) Counts of M,X-class flares as a function of the absolute relative angle between Mars and Earth in the years 1977-2019, for two different ranges of the absolute relative angle between Jupiter and Earth: ±5º around 120º (blue) and ±5º around 90º (red). (b) Same, only for X-class flares.

For either the presence or the absence of flares, Fig.6 lists the correlation values between the distributions for different pairs of planets. For a more accurate comparison to the days without flares, one entry per day is also considered for flares, in addition to the full sample with an arbitrary number of flares per day. Fig.6a shows the correlations for the full sample of flares, Fig.6b for individual days with flares, and Fig.6c for days without flares.

In these plots, empty cells correspond to zero. The standard deviation of the values for the days without flares is 0.29, and non-empty cells in Fig.6b exceed it by at least three standard deviations and in most cases notably more.

In summary, this Section presented a number of qualitative and quantitative indications for a non-random relation between the appearance of solar flares and the relative positions of the five innermost planets. The indications are mostly based on the distributions of flare counts as a function of the relative planetary angles, and seem to persist for combinations of more than two planets, and probably across different cycles and in the statistics of X-class flares. A few indicative plots were shown but the discussion applies to all relative angles between the five innermost planets. It could be added here that if either the C-class flares are included in the counts, or the total daily intensity is used instead of the counts, the results do not change qualitatively.

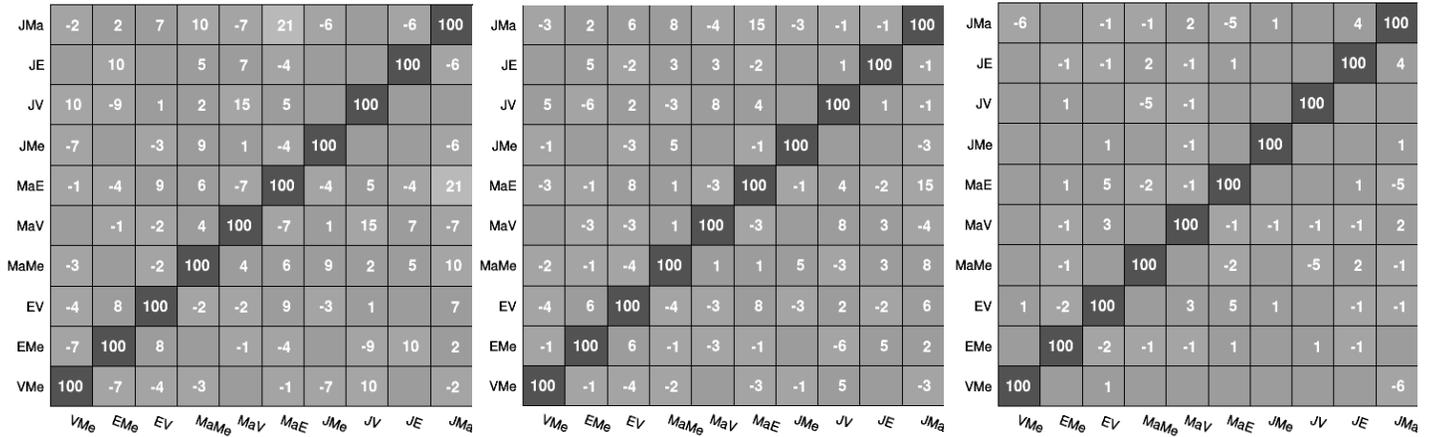

Figure 6. Correlations between distributions with either presence or absence of M,X-class flares as functions of the relative angle between pairs of the five innermost planets in the years 1977-2019. The names of the variables denote the first letters of each pair of planets. (a) Correlations for the full sample of flares, (b) for individual days with flares, (c) for individual days without flares.

## 3. Planetary alignments – The July alignment

Given the indications for a relation between solar flares and the relative planetary positions, it is reasonably expected that the alignment of several planets could play an enhanced role. For this reason, the case where five out of the six innermost planets align is examined.

Unlike in the previous Section, Saturn is now included. Previously, the change in the relative angles with respect to Saturn would be made meaningless by its slow motion, but alignments concern a specific configuration of the relative positions.

The definition of alignment will be having at least four out of the six innermost planets lie within ±15º around the line connecting the Sun and either Saturn or Jupiter (i.e. with respect to only one of the two for each given configuration). The alignment can take place on either side of the Sun; this equivalent treatment of conjunction and opposition is justified by its role in the planetary-based model of the solar cycle[1]. Finally, the alignment configuration should last for at least five days. This requirement is empirical and points both to a tighter alignment and probably to the need for an "effective time".

Notably, this configuration occurred just before the Carrington event (1859/09/01). In the following years and up to 1976 it occurred 20 times, but unfortunately the lack of continuous flares records does not allow further checks for that time length.

In the years 1977-2019 alignment occurred four times. The characteristics of these four occurrences and the accompanying solar activity are found in Table 1. Overall there are hints for increased activity around the alignments dates, but it is obvious that no statistical conclusions can be reached.

However, it is worth noting that the next alignment will occur on 2020/06/30-07/11 (Table 2). Currently we are at the beginning of a new cycle so activity is expected to be very low, but if on those dates there is an increased number of sunspots, then the Sun could be watched for flares as well. (As a side note, the random forest method described in the next Section classifies several days before and after these alignment dates as probable days with flares.)

| middle date | duration of alignment (days) | stage in cycle | M(/X) flare count | sunspot count | monthly sunspot average | comments |
|---|---|---|---|---|---|---|
| 19810401 | 20 | close to peak of 21 | 43/6 | 181-287 | 135, 156 | significant activity; not very strong flares |
| 19900108 | 7 | strong part of 22 | 6 | 160-210 | 240 | moderate activity |
| 20101012 | 10 | start of 24 (9 months in) | 1 | 0-57 | 23 | highest 5-day spot count for the cycle's first year |
| 20110602 | 6 | start of 24 (16 months in) | 0 | 80-135 | 37 | highest 5-day spot count before the cycle reaches 100 daily spots regularly |

Note: The daily sunspot count around the Carrington event was ~300, for monthly averages of 203 and 201.

Table 1. Features of the four occurrences of planetary alignments in the years 1977-2019.

| Mercury | Venus | Earth | Mars | Jupiter | Saturn |
|---|---|---|---|---|---|
| 288 | 301 | 282 | 317 | 291 | 298 |

Table 2. Planetary heliocentric ecliptic longitudes on 2020/07/04 (in degrees).

## 4. Classification with random forests

An ensemble of decision trees was trained on the daily planetary angles to learn how to differentiate between days with and without flares of M,X classes. The results hint at the predictive power of this approach, even though it does not make use of any solar observables.

The ensemble contained 200 trees, optimized through gradient boosting. The sample consists of the full record of 6,855 flares, appearing on 3,285 days, and of 12,420 days without flares. The analysis was performed with the TMVA toolkit[10].

The variables are the ten relative angles formed by the five innermost planets, and a variable reflecting the strength of the solar cycle on each day, since the same planetary configurations are expected to have accordingly different outcomes. This variable is derived from the model of the solar cycle based on the relative motion of Jupiter and Saturn[1]; it consists of the daily value of the model subtracted from the maximum value for the given cycle.

In the simplest approach, the validation sample is randomly selected. Fig.7a shows the classification output for the training and validation samples, and Fig.7b shows the corresponding ROC curve; the validation sample consists of 200 events of each category. "Signal" and "background" denote the days with and without flares respectively.

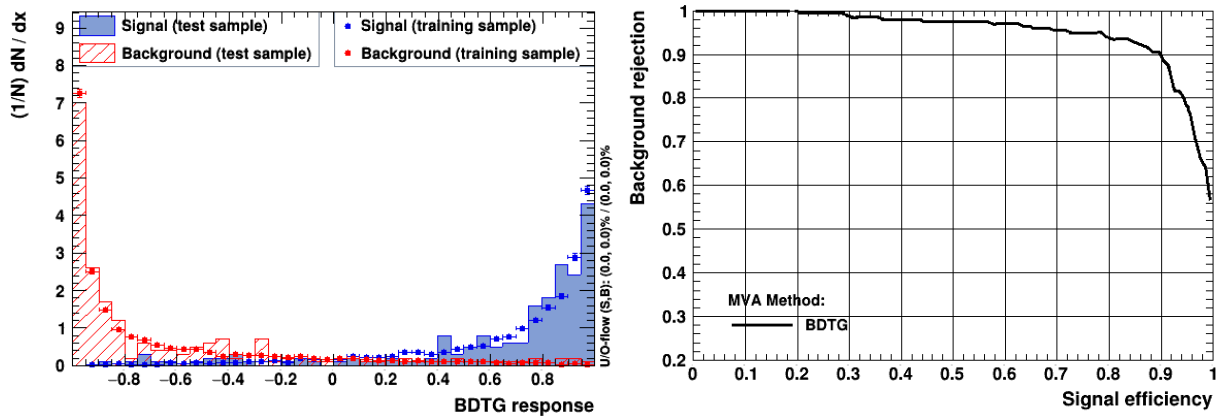

Figure 7. Random forest training on the values of relative planetary angles for classification of days with ("signal") and without ("background") M,X-class flares in the years 1977-2019, and validation on randomly selected events. (a) Classification output. (b) ROC curve.

Although the classification of the random validation sample looks strong, it is biased by the neighboring data points, i.e. almost all days in time ranges with overall high solar activity are classified as signal and vice versa (Finley effect).

To circumvent this bias, the validation was performed on a sliding window of one year, while the training used the rest of the sample. Fig.8a shows the number of correctly classified events in each category, separately for each year; the selection cut on the classification outputs was set to zero for all years. The plot enables the assessment for a variety of different "populations". However, the relative number of correctly classified events in each category can vary according to the value of the selection cut on the classification outputs; instead, Fig.8b shows the sum of the correctly classified events in both categories, which is expected to remain more stable.

The Finley effect is visible in these results, however a tendency for more sophisticated classification is also present. An example might be the year 2017, which was marked by an increase in activity in late summer. The dates for ten out of the 15 days with flares of that year are correctly classified, with nine fake positives. In total for all years, 62% of days were correctly classified.

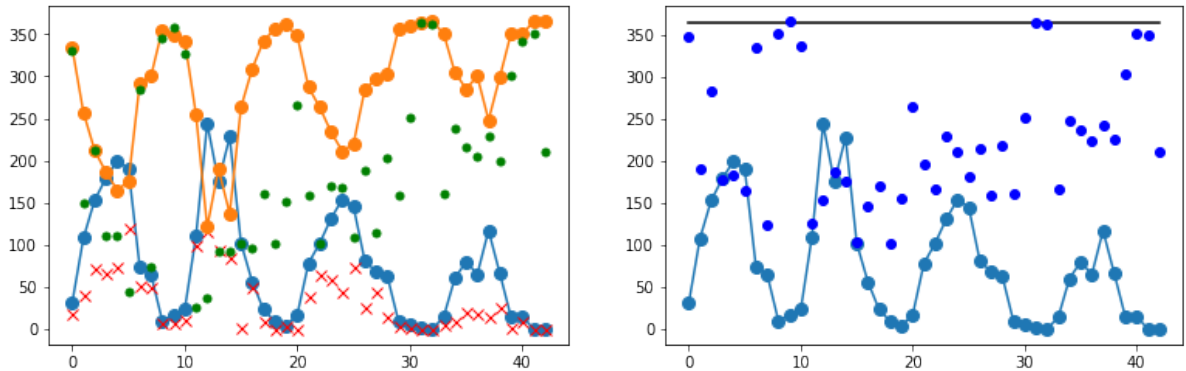

Figure 8. Validation of random forest training on the values of relative planetary angles for classification of days with and without M,X-class flares, with the validation performed individually for each year in 1977-2019. (a) Blue histogram: Number of days with flares in each year. Orange histogram: Number of days without flares. Red crosses: Number of correctly classified days with flares. Green dots: Number of correctly classified days without flares. (b) Blue histogram: Number of days with flares. Blue dots: Sum of the number of correctly classified days with and without flares. A visual guide is drawn at 365.

This preliminary application of machine learning reinforces the conjecture for a relation between planetary configurations and the presence of flares on short timescales, since the training did not make use of any solar observable. Also, it shows that there is promise for the use of planetary positions in the forecasting of solar activity.

## 5. Conclusions

This article presented a number of indications for a non-random relation between the appearance of solar flares of classes M and X, and the relative motion of the five innermost planets. The indications were based mostly on the distributions of flares in the years 1977-2019 as a function of the relative heliocentric longitude between pairs of planets. The special case of alignment of several planets was also examined, given the fact that the next such alignment takes place in the following weeks.

The conjecture of such a relation was reinforced by a preliminary application in forecasting. A random forest trained on the planetary angles provides promising classification for the presence and absence of flares, without the use of solar observables.

At present no proposal is made about underlying physical mechanisms, but the phenomenological results justify a deeper look. Physical understanding, and more comprehensive tests, will probably come from the inclusion of more solar effects.